\documentclass[%
 reprint,
 amsmath,amssymb,
prd,
]{revtex4-2}
\usepackage[colorlinks=true,citecolor=blue,filecolor=blue,linkcolor=blue,urlcolor=blue,pdftex]{hyperref}
\usepackage{graphicx}
\usepackage{dcolumn}
\usepackage{bm}

\usepackage[dvipsnames]{xcolor}
\usepackage{orcidlink}

\def\equationautorefname~#1\null{Equation~(#1)\null}

\newcommand{\unit}[1]{\, \mathrm{#1}}
\begin{document}


\title{Fast estimation of the look-elsewhere effect using Gaussian random fields}

\author{Juehang Qin, \orcidlink{0000-0001-8228-8949}}
\email{qin106@purdue.edu}
\author{Rafael F. Lang, \orcidlink{0000-0001-7594-2746}}%
\affiliation{%
 Department of Physics and Astronomy, Purdue University, West Lafayette, IN 47907, USA
}%
 

\begin{abstract}
We discuss the use of Gaussian random fields to estimate the look-elsewhere effect correction. We show that Gaussian random fields can be used to model the null-hypothesis significance maps from a large set of statistical problems commonly encountered in physics, such as template matching and likelihood ratio tests. Some specific examples are searches for dark matter using pixel arrays, searches for astronomical transients, and searches for fast-radio bursts. Gaussian random fields can be sampled efficiently in the frequency domain, and the excursion probability can be fitted with these samples to extend any estimation of the look-elsewhere effect to lower $p$-values. In addition, in cases where the Gaussian random field is stationary and the parameter space is Euclidean, the look-elsewhere effect correction can be computed analytically. We demonstrate these methods 
using two example template matching problems. Finally, we apply these methods to estimate the trial factor of a \(4^3\) accelerometer array for the detection of dark matter tracks in the Windchime project. When a global significance of \(3\sigma\) is required, the estimated trial factor for such an accelerometer array is \(10^{14}\) for a one-second search, and \(10^{22}\) for a one-year search.
\end{abstract}

\maketitle


\section{Introduction}

In hypothesis testing problems with composite hypotheses, the correct frequentist 
$p$-value might not be the same as the $p$-value one would compute for fixed values of the composite hypothesis parameters~\cite{ParticleDataGroup:2022pth}. This is referred to as the look-elsewhere effect. The correct $p$-value given composite hypotheses is often termed the global $p$-value, whereas the $p$-value computed with fixed parameters is termed the local $p$-value. The look-elsewhere effect correction is often parameterized by a trial factor, which is the ratio between the local and global $p$-values~\cite{Gross:2010qma}. A simple approach to finding the trial factor would be to run a large number of null-hypothesis Monte-Carlo simulations, often using simplified or ``toy" models that retain the relevant statistical properties.  An example of this approach can be found in~\cite{XENON:2020rca}.

However, data-analysis and inference of modern experiments can be extremely computationally-intensive, requiring dedicated computational infrastructure. Even then, with complex and high-dimensional problems, the use of toy Monte Carlo simulations can be computationally too expensive. This makes estimation of trial factors for the purposes of sensitivity projections for future experiments difficult. Thus, the use of Gaussian random fields to directly generate null-hypothesis significance maps (termed `null significance maps') and estimate the trial factor can be very useful.

Gaussian random fields are random functions over a domain, where the values of every finite collection of points on the domain are described by a multivariate Gaussian distribution. In parameter estimation problems, the domain would be the parameter space as defined by the relevant parameters, such as a finite two-dimensional Euclidean space for a search for a transient in an image. Such fields can be viewed as a higher-dimensional generalization of Gaussian processes, commonly used in Gaussian process regression~\cite{Carleo:2019ptp}. Gaussian random fields are used for the estimation of the look-elsewhere effect in neuroimaging~\cite{Worsley:doi:10.1038/jcbfm.1992.127}, and for the modelling of the matter distribution in the universe~\cite{Bardeen:1985tr}. Work exists regarding the use of Gaussian random fields for the estimation of the look-elsewhere effect in physics~\cite{Vitells:2011APh....35..230V, Ananiev:2022kwm}. The use of Gaussian random fields to estimate look-elsewhere effect corrections relies on computation of the excursion probability, which is the probability for samples of a Gaussian random field to exceed a given significance level.

In this paper, we detail techniques to use Gaussian random fields for the estimation of look-elsewhere effect corrections, with a focus on problems with underlying Gaussian random variables. We expand upon the work presented in~\cite{Vitells:2011APh....35..230V} by including a way to estimate the look-elsewhere effect correction analytically in Euclidean parameter spaces without requiring computations of the Euler characteristic from data or simulation. Our work also differs from~\cite{Ananiev:2022kwm}, which focuses on how to compute a covariance matrix using specially constructed mock datasets, termed Asimov datasets~\cite{Cowan:2010js}, and then estimating the trial factor from the covariance matrix as a 1D Gaussian process. Our work instead uses higher-dimensional Gaussian fields directly, hence allowing for direct computation of the trial factor in arbitrarily high dimensional parameter spaces if the spectral moments of the Gaussian random field can be directly computed or expressed. These methods are complementary, as our method does not require the computation of the entire covariance matrix at high resolution, which might be difficult for high-dimensional parameter spaces, whereas~\cite{Ananiev:2022kwm} does not assume stationarity in the Gaussian process, which is useful for more complex statistical problems. We also include methods to sample stationary and Euclidean Gaussian random fields in the frequency domain to reduce computational expense.

In~\autoref{sec:maths}, we discuss the classes of problems that can be modelled by Gaussian random fields, overview the spectral method for efficient sampling of Gaussian random fields, and introduce an analytic approximation for the excursion probability.
We then demonstrate these methods using a 2-dimensional template-matching problem with a Gaussian kernel, and a 1-dimensional template-matching problem with a non-Gaussian kernel, in~\autoref{sec:2D_demo} and~\autoref{sec:1D_demo} respectively. Then, in \autoref{sec:peak_search}, we demonstrate how these methods can be applied to a likelihood ratio test, where a check for asymptotic behaviour is needed. Finally, in~\autoref{sec:windchime_demo}, we use these methods to estimate the trial factor when searching for dark matter tracks using an array of accelerometers.

\section{Statistical underpinnings of method}\label{sec:maths}

This section is split into 3 parts. First, we explore why Gaussian random fields correctly model the significance maps of a large set of problems in~\autoref{ssec:random_field_model_explanation}. Second, we discuss how to sample Gaussian random fields efficiently in~\autoref{ssec:stationary_field_sampling}. Finally, in~\autoref{ssec:analytic_approx_excursion_prob} we describe a fitting procedure that can be used to extend this method to small $p$-values using an analytic approximation of the excursion probability of a Gaussian random field, and a way to directly estimate the excursion probability using the Euler characteristic when certain conditions that will be elaborated upon are met.

\subsection{Why Gaussian random fields can model a large set of problems}\label{ssec:random_field_model_explanation}

\begin{figure}[htp]
 \centering
    \includegraphics[width=\columnwidth]{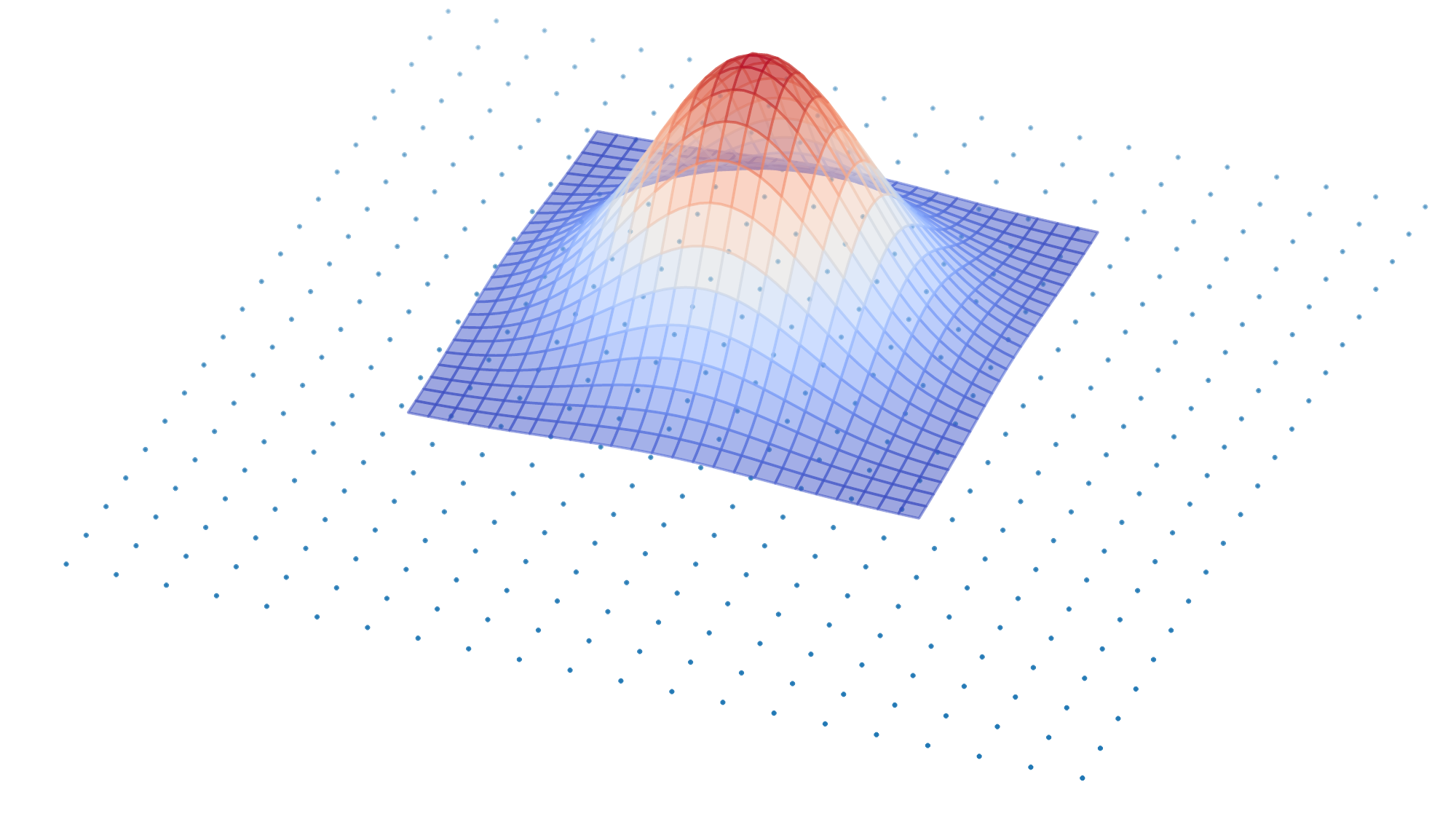}%
    \caption{Diagram depicting a template matching search for an excess over a grid of random variables. The grid of random variables is indicated by the blue points, and the template used to search for an excess is shown as the colored wire-frame distribution.}\label{fig:template_matching_2d}
\end{figure}

Let us consider a statistical problem where one searches for a fluctuation over a finely-spaced set of Gaussian random variables distributed in a parameter space. One example of this could be a template matching search for a transient over a regular grid of CCD pixels with Gaussian noise, as shown in Fig.~\ref{fig:template_matching_2d}. Such a setup might be encountered in searches for astronomical transients~\cite{Andreoni:2019jlb, LSST:2022kad}.

While a 2D grid with a simple template that is symmetric and does not vary with position is depicted in Fig.~\ref{fig:template_matching_2d}, this is for ease of illustration, and these assumptions are not made in the following argument except where noted. We can see that at each possible template position, the resultant signal strength recovered is a weighted sum of Gaussian random variables, where the weights correspond to the template amplitude at a given random variable. The significance map formed using such a template thus corresponds to the formal definition of a random field \cite{Adler:2007-06-12} where each point is Gaussian-distributed, and every collection of points represent a multivariate Gaussian distribution.

In addition, given independent underlying random variables, the covariance between two points can be computed from the template directly. Consider two points in the parameter space, \((\mathbf{x}_0, \mathbf{x}_1)\). At each point, the random variable in the case of a signal-free dataset is given by:
\begin{equation}\label{eq:Y_i}
Y_i = \sum_j \alpha_{i,j} X_j    
\end{equation}
where \(X_j\) are the underlying finely spaced Gaussian random variables such as CCD pixels, and \(\alpha_{i,j}\) refers to the template value at each underlying random variable for template \(i\). the expected value of $X_j$, \(\mathbb{E}(X_j)=0\), is taken without loss of generality, as the mean value can be subtracted. As such, the covariance would be given by:
\begin{equation}\label{eq:covariance_two_point}
\begin{split}
K(\mathbf{x_0}, \mathbf{x_1}) &= \mathrm{cov}\left(\sum_j \alpha_{0,j} X_j, \sum_k \alpha_{1,k} X_k\right)\\
&= \mathbb{E}\left(\left(\sum_j \alpha_{0,j} X_j\right)\left(\sum_k \alpha_{1,k} X_k\right)\right)
\end{split}
\end{equation}

This can be further simplified if the underlying Gaussian random variables are independent, as then \(\mathbb{E}(X_i X_j) = 0\) for \(i \neq j\). The covariance can then be computed:
\begin{equation}\label{eq:covariance_two_point_indep}
\begin{split}
K(\mathbf{x_0}, \mathbf{x_1}) &= \mathbb{E}\left(\left(\sum_j \alpha_{0,j} X_j\right)\left(\sum_k \alpha_{1,k} X_k\right)\right)\\
&= \mathbb{E}\left(\sum_j \alpha_{0,j}\alpha_{1,j} X_j^2\right)\\
&= \sum_j \alpha_{0,j}\alpha_{1,j} \sigma_j^2
\end{split}
\end{equation}

We can thus see that in the case of template matching with underlying Gaussian random variables, the significance map is modelled by a Gaussian random field and the covariance function can be directly calculated based on the template as well as the measured properties of the underlying random variables. In addition, Gaussian random fields can also be used to model significance maps with underlying Gaussian random variables generated from likelihood ratio tests. This is because the log likelihood-ratio is the sum of squared residuals, normalized by the standard deviation, as shown in \autoref{eq:gaussian_llr}.
\begin{equation}\label{eq:gaussian_llr}
\begin{split}
\Lambda &= \log\left(\frac{\hat{\mathcal{L}}}{\mathcal{L}_0}\right)\\
&= \sum_i \frac{-\left(X_i - \mu_i\right)^2}{2\sigma} - \sum_i \frac{-\left(X_i\right)^2}{2\sigma}\\
&= \sum_i \frac{2\mu_i X_i - \mu_i^2}{2\sigma}
\end{split}
\end{equation}

Some other problems that do not use underlying Gaussian random variables can still be represented approximately by Gaussian random fields in some circumstances. For example, if a likelihood ratio test is used, the distribution of the test statistic asymptotically approaches a \(\chi^2\) distribution due to Wilks' theorem~\cite{Algeri:2019lah} when the relevant conditions, detailed in \cite{Algeri:2019lah}, are satisfied. In such a situation, a signal-free significance map over a parameter space would represent a \(\chi^2\) random field \cite{Vitells:2011APh....35..230V}. This differs from a Gaussian random field. However, because a \(\chi^2\) random variable is defined as the square of a Gaussian random variable, a \(\chi^2\) random field can be sampled by sampling a Gaussian random field with the correct covariance, then squaring it. The excursion probability of such a random field is thus double the one-sided excursion probability described in \autoref{ssec:analytic_approx_excursion_prob}. The fact that a \(\chi^2\) random field can be represented using Gaussian random field was noted by Ananiev and Read in \cite{Ananiev:2022kwm}.

Due to Lindeberg and Lyapunov's central limit theorems,~\cite{Lindeberg1922-ir, Billingsley_Patrick2012-01-20}, this even holds for non-Gaussian underlying random variables such as Poisson noise, as long as Lindeberg's condition or Lyapunov's condition are satisfied \cite{Billingsley_Patrick2012-01-20}. More generally, our results are applicable whenever the underlying random variables are independent and sufficiently finely distributed to give a large sample size, as long as the criteria for the central limit theorems are met. Template matching problems can then be approximated by Gaussian random fields even if the underlying random variables are non-Gaussian. While a detailed discussion of these conditions is beyond the scope of this paper, verifying the Gaussianity of a given significance map numerically or constraining it using the Berry-Esseen theorem \cite{Tyurin:2009arXiv0912.0726T} might be sufficient for practical purposes. Taken together, the methods we discuss in this paper are applicable to a large variety of statistical problems commonly encountered in experimental physics.

It should be noted that such methods based on Gaussian random fields are limited in usefulness in cases where the noise sources are not Gaussian in nature and the central limit theorem does not apply. Moreover, these methods do not account for other sources of false positives such as experimental backgrounds.

\subsection{Efficient spectral sampling of stationary Gaussian random fields}\label{ssec:stationary_field_sampling}

While for sufficiently complex problems, Gaussian random fields might be easier to sample than toy Monte Carlo-based methods, even this might still be too computationally intensive. For example, if we consider template-matching search in a flat 2-dimensional parameter space with \(10^2\) bins per dimension, there would be \(10^4\) points that need to be correlated with each other, resulting in a covariance matrix with \(10^8\) entries. We can see that with higher dimensional problems, populating such a covariance matrix (which is needed for naive sampling of Gaussian random fields) quickly becomes intractable.

A review of efficient methods for the sampling of Gaussian random fields can be found in \cite{liu2019advances}. In sections~\ref{sec:2D_demo} and \ref{sec:windchime_demo}, we use the spectral method as described in \cite{liu2019advances} to efficiently sample from Gaussian random fields. This method of generating samples from a Gaussian random field requires the field to be weakly stationary, such that a covariance function can be described by a function of the displacement between two points. In that case,
\begin{equation} \label{eq:stationary_covariance}
    K(\mathbf{x_0}, \mathbf{x_1}) \equiv K_s(\mathbf{x_0} - \mathbf{x_1})
\end{equation}
where \(K_s(\mathbf{s})\) is the autocorrelation function for a random field of zero mean. For a field with unity variance, which is typical for a significance map that is scaled to represent the signal-to-noise ratio, \(K_s(\mathbf{s})\) can be scaled to have a maximum value of one. This case is considered without loss of generality, as one can scale any stationary Gaussian random field to have a variance of unity. The Fourier transform of the autocorrelation function of a weakly stationary process is the power spectral density (PSD) due to the Wiener-Khinchin theorem \cite{Vetterli:2014-10-20, Jaynes_E_T_2003-04-10}.

The spectral method for sampling Gaussian random fields makes use of this Fourier transform pair. Due to the Wiener-Khinchin theorem, instead of sampling a stationary Gaussian random field with a given autocorrelation function directly, it can be sampled in the frequency domain with the correct PSD as given by the Fourier transform of the autocorrelation function. This can be done by multiplying the Fourier transform of white noise by the square root of the PSD, then performing an inverse Fourier transform to return the sample to the relevant parameter space. We can thus sample stationary Gaussian random fields in high-dimensional parameter space without having to populate extremely large covariance matrices. For the remainder of text, we will refer to this method for sampling stationary Gaussian random fields as the spectral method.

\subsection{Analytic approximation of excursion probability}\label{ssec:analytic_approx_excursion_prob}

Even when sampling Gaussian random fields with the spectral method, in high-dimensional parameter spaces, sampling can still be prohibitively expensive due to the curse of dimensionality. In this scenario, the look-elsewhere effect correction can be extended to lower $p$-values than otherwise feasible due to computational requirements using an analytic approximation of the excursion probability. For a random field \(\left\{ f(t):t \in M \right\}\), the excursion probability over a confidence level \(u\) is defined as:
\begin{equation} \label{eq:excursion_probability}
    p_{\mathrm{excur}} = \mathbb{P}\left\{\sup_{t \in M} f(t) \geq u \right \}
\end{equation}
It can be seen that the excursion probability represents the probability that any point on a random field exceeds \(u\). The excursion probability for a smooth Gaussian random field on a locally convex space is given by~\cite{Adler:2007-06-12}:
\begin{equation} \label{eq:excursion_probability_approx}
\begin{split}
    p_{\mathrm{excur}} = &C_0 \Phi \left(\frac{u}{\sigma}\right) + u^N e^{-u^2/2\sigma^2} \sum^N_{j=1} C_j u^{-j} + \\
    &\mathcal{O}\left(e^{-\alpha u^2/2\sigma^2}\right)
\end{split}
\end{equation}
where \(\Phi(u)\) is the Gaussian tail distribution, \(N = \mathrm{dim}(M)\), \(\sigma\) is the standard deviation of the Gaussian random field, and \(\alpha>1\) is a constant describing the exponential suppression of the error term. We can see that \autoref{eq:excursion_probability_approx} contains a number of constants \((C_i)\). While these constants can be computed directly for some cases, this is sometimes non-trivial \cite{Adler:2007-06-12}. In these cases, \autoref{eq:excursion_probability_approx} can be used to fit the excursion probability directly using a set of samples of signal-free significance maps. This, combined with the spectral method of sampling Gaussian random fields shown in \autoref{ssec:stationary_field_sampling}, greatly reduces the computational cost of computing the look-elsewhere effect correction. It should be noted that \autoref{eq:excursion_probability_approx} is based on approximating the excursion probability using the mean Euler characteristic of an excursion set~\cite{Adler:2007-06-12}. Thus, for particularly computationally challenging problems, methods from \cite{Vitells:2011APh....35..230V} for estimating the Euler characteristic can further reduce the number of samples that are needed to derive the look-elsewhere effect correction.

The Euler characteristic $\varphi$ of an excursion set $A_u$ can be interpreted as the number of isolated regions in an excursion set where the Gaussian random field $f(t)$ exceeds some significance level $u$ $(f(t) > u)$, minus the number of holes in these isolated regions~\cite{Worsley:doi:10.1038/jcbfm.1992.127}; a precise mathematical definition can be found in~\cite{Adler:2007-06-12}. If the Gaussian random field is on a rectangular Euclidean space $M = \prod^{N}_i=1 [0, M_i]$, the mean Euler characteristic can be expressed in terms of the dimensions of the Euclidean space and the derivatives of the random field~\cite{Adler:2007-06-12}:
\begin{equation}\label{eq:euler_characteristic}
\begin{split}
    \mathbb{E}\left(\varphi(A_u)\right) =& e^{-{u^2/2\sigma^2}}+\\
    &\sum^N_{k=1}\sum_{J\in\mathcal{O}_k}\frac{|J||\Lambda_J|^{1/2}}{(2\pi)^{(k+1)/2}\sigma^k}H_{k-1}\left(\frac{u}{\sigma}\right)+\\
    &\Phi\left(\frac{u}{\sigma}\right),
\end{split}
\end{equation}
where $\mathcal{O}_k$ refers to the set of $k$-dimensional faces of $M$ that contain the origin, and $|J|$ refers to the volume of the faces. Iterating over the $k$-dimensional faces that contain the origin can also be interpreted as iterating over $k$-dimensional slices of the space $M$. $\Lambda_J$ is a $k\times k$ matrix containing the spectral moments of the Gaussian random field~\cite{Adler:2007-06-12}. These can be computed explicitly using the second derivative of the covariance function~\cite{Adler:2007-06-12}:
\begin{equation}\label{eq:lambda_ij}
    \lambda_{ij} = -\left.\frac{\partial^2}{\partial d^i \partial d^j} K_s(\mathbf{s})\right|_{\mathbf{s} = \vec{0}}
\end{equation}
where $\lambda_{ij}$ refers to the element of $\Lambda_J$ in the $i^\mathrm{th}$ row and the $j^\mathrm{th}$ column, $d^i$ refers to the $i^\mathrm{th}$ element of the $\mathbf{s}$ vector, and $K_s(\mathbf{s})$ is the covariance function as defined in~\autoref{eq:stationary_covariance}.

For a problem where the point response function is known, this can be computed directly using the following integral~\cite{Worsley:doi:10.1038/jcbfm.1992.127}:
\begin{equation}\label{eq:lambda_prf}
    \Lambda_J = \left . \int\frac{\partial \alpha_s(\mathbf{s})}{\partial \mathbf{s}}\frac{\partial \alpha_s(\mathbf{s})}{\partial \mathbf{s}^T} d\mathbf{s}\,\middle/\,\int\alpha_s(\mathbf{s})^2 d\mathbf{s}\right.
\end{equation}
where $\alpha_s(\mathbf{s})$ is the point response function, and $\mathbf{s}^T$ denotes the transpose of the $\mathbf{s}$ vector. This is equivalent to the template function used in~\autoref{eq:pos_independent_covariance} where the field is stationary, such that $\alpha_s(\mathbf{x} - \vec{x}) = \alpha(\mathbf{x}, \vec{x})$. For the case of a Gaussian kernel with covariance matrix $\Sigma$, we obtain $\lambda_J = \Sigma^{-1}/2$~\cite{Worsley:doi:10.1038/jcbfm.1992.127}.

For problems with stationary fields and Euclidean parameter spaces where~\autoref{eq:euler_characteristic} applies, this equation can be interpreted as an approximation of $p_{\mathrm{excur}}$. Thus, \autoref{eq:euler_characteristic} can be combined with either \autoref{eq:lambda_ij} or \autoref{eq:lambda_prf} to produce an estimate of $p_{\mathrm{excur}}$.

\section{Demonstration with a 2D toy problem}\label{sec:2D_demo}

The ideas introduced in \autoref{sec:maths} can be first demonstrated using a search in a 2D parameter space. Here, we will model a 2D template matching search using a 2D Gaussian random field. 2D Gaussian random fields can be used to model the look-elsewhere correction for various experiments, such as searches for dark matter using pixel detectors \cite{DAMIC:2013bio, DAMIC:2021crr} and searches for astronomical transients \cite{Andreoni:2019jlb, LSST:2022kad}.

\begin{figure}[htp]
 \centering
    \includegraphics[width=\columnwidth]{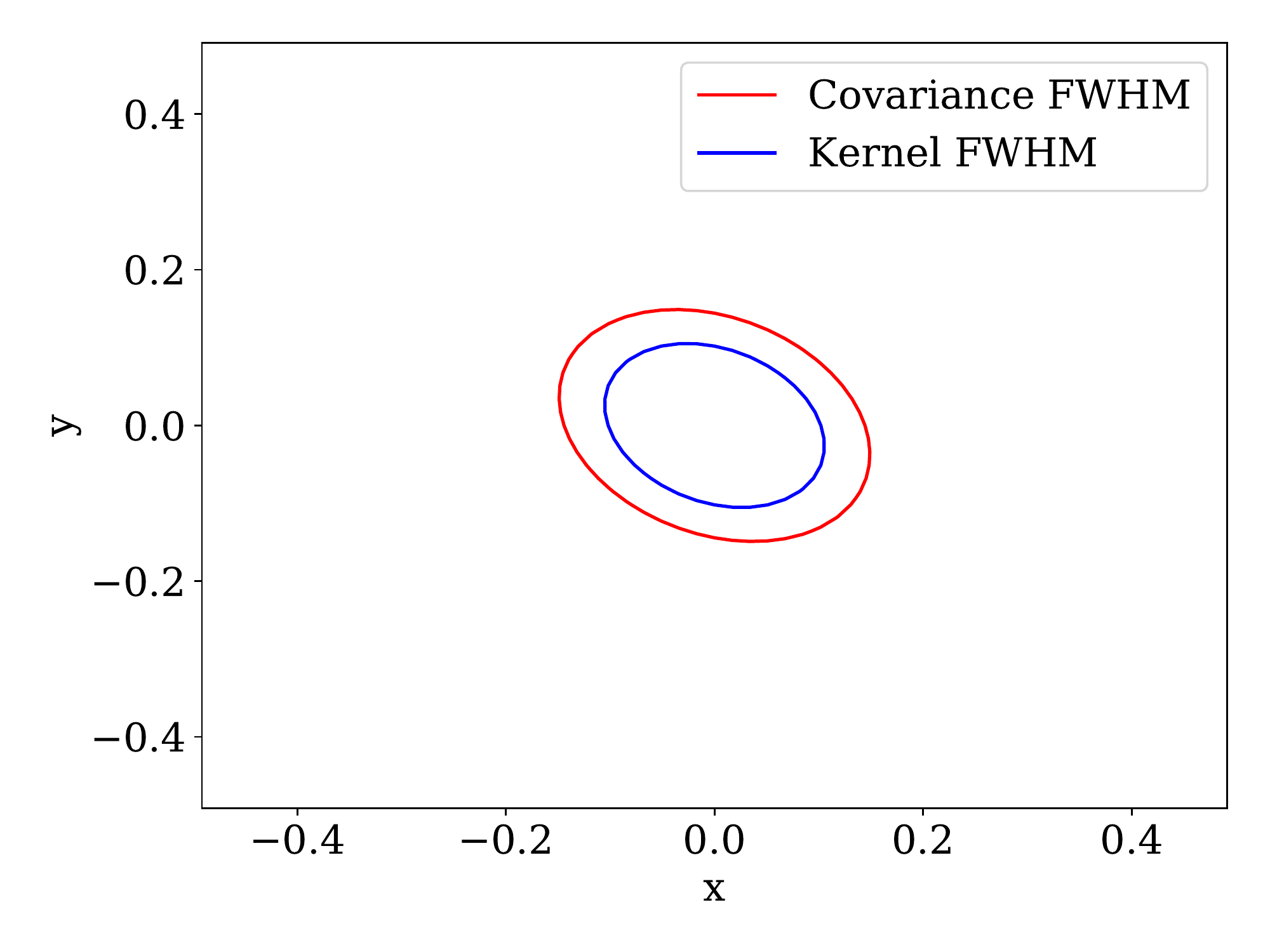}%
    \caption{2-dimensional toy problem template matching kernel and covariance function. The ellipse corresponding to the full width at half maximum (FWHM) of the covariance function is \(\sqrt{2}\) bigger than that of the kernel in linear dimension.}\label{fig:2D_kernel}
\end{figure}

The expected signal shape is chosen to be a Gaussian kernel for this problem; hence, the template matching kernel can also be modelled using a Gaussian kernel. In such a situation, the normalized covariance function (the correlation function) is also a Gaussian kernel with double the covariance matrix and \(\sqrt{2}\) the linear dimensions, as shown in \autoref{fig:2D_kernel}.

This can be seen by expanding \autoref{eq:covariance_two_point_indep} in the case of a position-independent kernel:
\begin{equation} \label{eq:pos_independent_covariance}
    K(\mathbf{x_0}, \mathbf{x_1}) = \sum_j \alpha(\mathbf{x_0}, \vec{x}_j) \alpha(\mathbf{x_1}, \vec{x}_j) \sigma_j^2
\end{equation}
where \(\alpha(\mathbf{x_0}, \vec{x}_j)\) is the template weight for a sample at \(\vec{x}_j\), and a template at \(\mathbf{x_0}\). For the case of a stationary field, \(\sigma_j\) is the same for all \(j\), so we call this simply \(\sigma\). Then, taking the template function to be a Gaussian with covariance matrix \(\Sigma\), we can apply the continuum approximation if each data samples takes up a volume of \(V_{\mathrm{s}}\). Thus, we arrive at:
\begin{equation} \label{eq:gauss_continuum_approx}
\begin{split}
K(\mathbf{x_0}, \mathbf{x_1}) =& \sigma^2 \sum_j \alpha(\mathbf{x_0}, \vec{x}_j) \alpha(\mathbf{x_1}, \vec{x}_j)\\
\approx& \frac{\sigma^2}{V_{s}} \int \alpha(\mathbf{x_0}, \vec{x}) \alpha(\mathbf{x_1}, \vec{x}) d\vec{x}\\
=& \frac{\sigma^2}{V_{s}} \int e^{\left(\mathbf{x_0} - \vec{x}\right)^T \Sigma^{-1} \left(\mathbf{x_0} - \vec{x}\right)} e^{\left(\mathbf{x_1} - \vec{x}\right)^T \Sigma^{-1} \left(\mathbf{x_1} - \vec{x}\right)} d\vec{x}
\end{split}
\end{equation}

\begin{figure}[htp]
 \centering
    \includegraphics[width=\columnwidth]{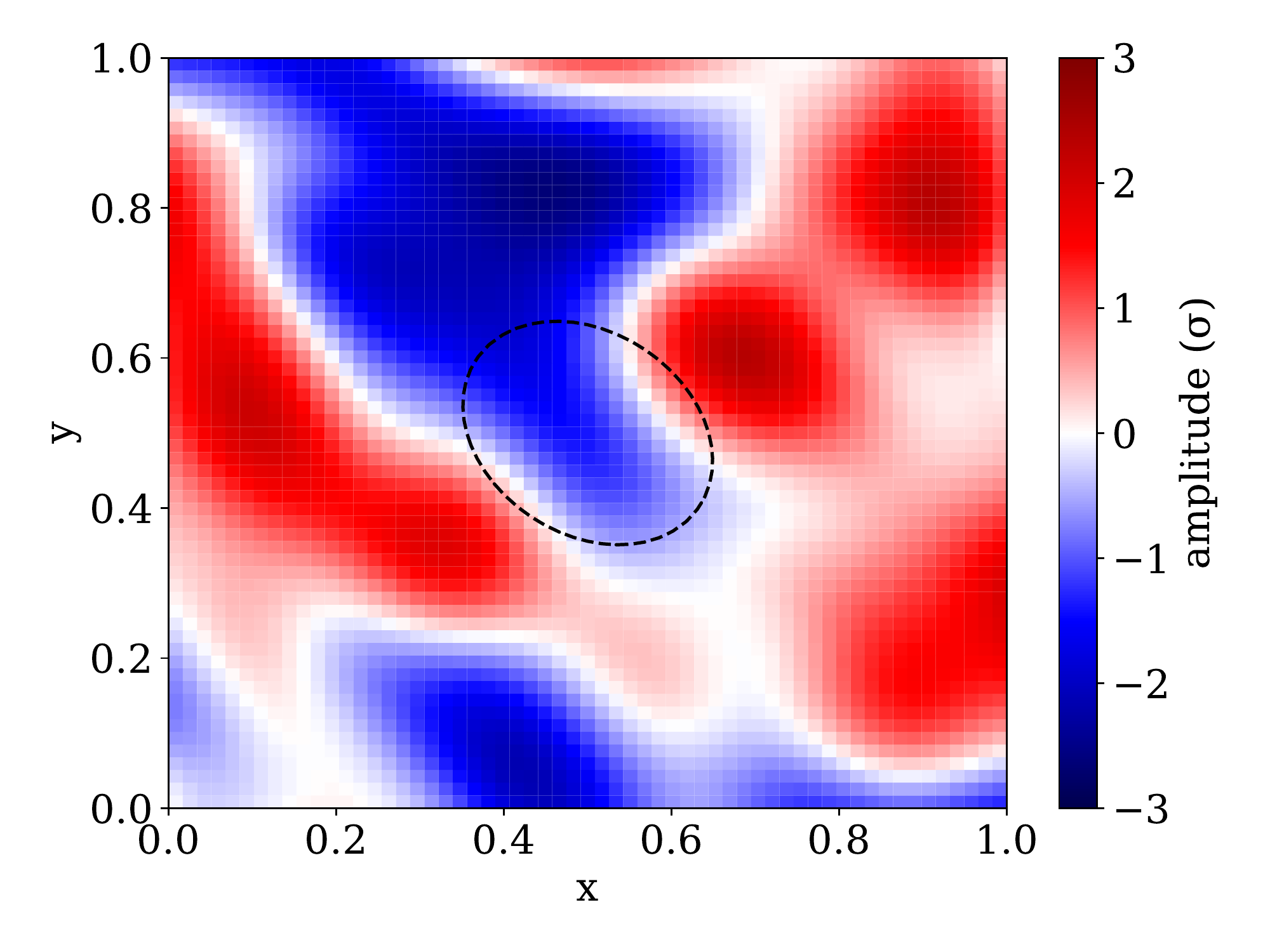}%
    \caption{Example of a single null random sample. The FWHM of the covariance function is overlaid as a black dashed ellipse for comparison.}\label{fig:2D_sample}
\end{figure}

We can see that this is linearly proportional to the convolution of two Gaussian distributions. It is a well known result that the convolution of Gaussian distributions is Gaussian, with a covariance matrix that is the sum of the covariance matrix of the individual Gaussian random variables\cite{Jaynes_E_T_2003-04-10}. Thus, if the templates in a search are Gaussian, the resulting Gaussian field from the template search has the covariance function:
\begin{equation}\label{eq:gauss_kernel_covariance}
    K(\mathbf{x_0}, \mathbf{x_1}) = K_s (\mathbf{x_0} - \mathbf{x_1}) \propto e^{\left(\mathbf{x_0} - \mathbf{x_1}\right)^T \Sigma'^{-1} \left(\mathbf{x_0} - \mathbf{x_1}\right)}
\end{equation}
where \(\Sigma'=2\Sigma\). One can then compute the normalization analytically using the convolution integral~\autoref{eq:gauss_continuum_approx}. In our case, we simply discard the normalization. This is not an issue as the goal is to produce a null significance map for the purpose of calibrating the look-elsewhere effect, and such significance maps are normalized to have unity variance so that the local significance is directly given by the significance map. A single sample from the Gaussian random field described by \autoref{eq:gauss_kernel_covariance} is shown in \autoref{fig:2D_sample}. The parameter space is divided into \(60\) bins in each dimension when generating this sample.

\begin{figure}[htp]
 \centering
    \includegraphics[width=\columnwidth]{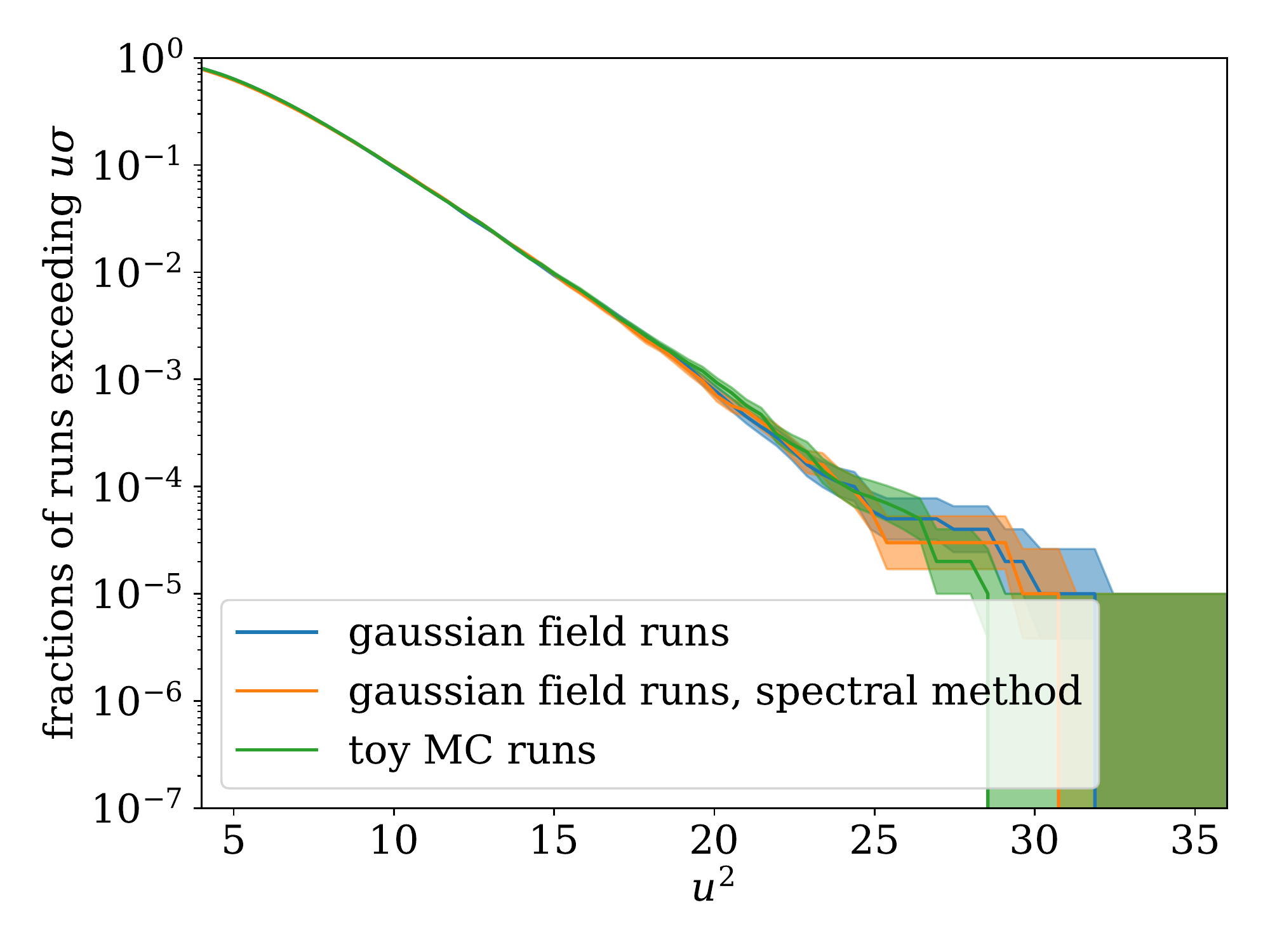}%
    \caption{The fraction of \(10^5\) null random samples showing false positives as a function of the significance threshold in units of \(\sigma^2\). We can see that the different methods to generate random fields produce global $p$-values that are in agreement.}\label{fig:2D_comparison}
\end{figure}

We can now compare samples generated using a traditional toy Monte Carlo and Gaussian field samples. This is shown in \autoref{fig:2D_comparison}. The Gaussian random field is sampled both using the spectral method described in \autoref{ssec:stationary_field_sampling} and a naive method, where we directly sample a large covariance matrix describing the covariance between every pair of points as a multivariate Gaussian.

\begin{figure}[htp]
 \centering
    \includegraphics[width=\columnwidth]{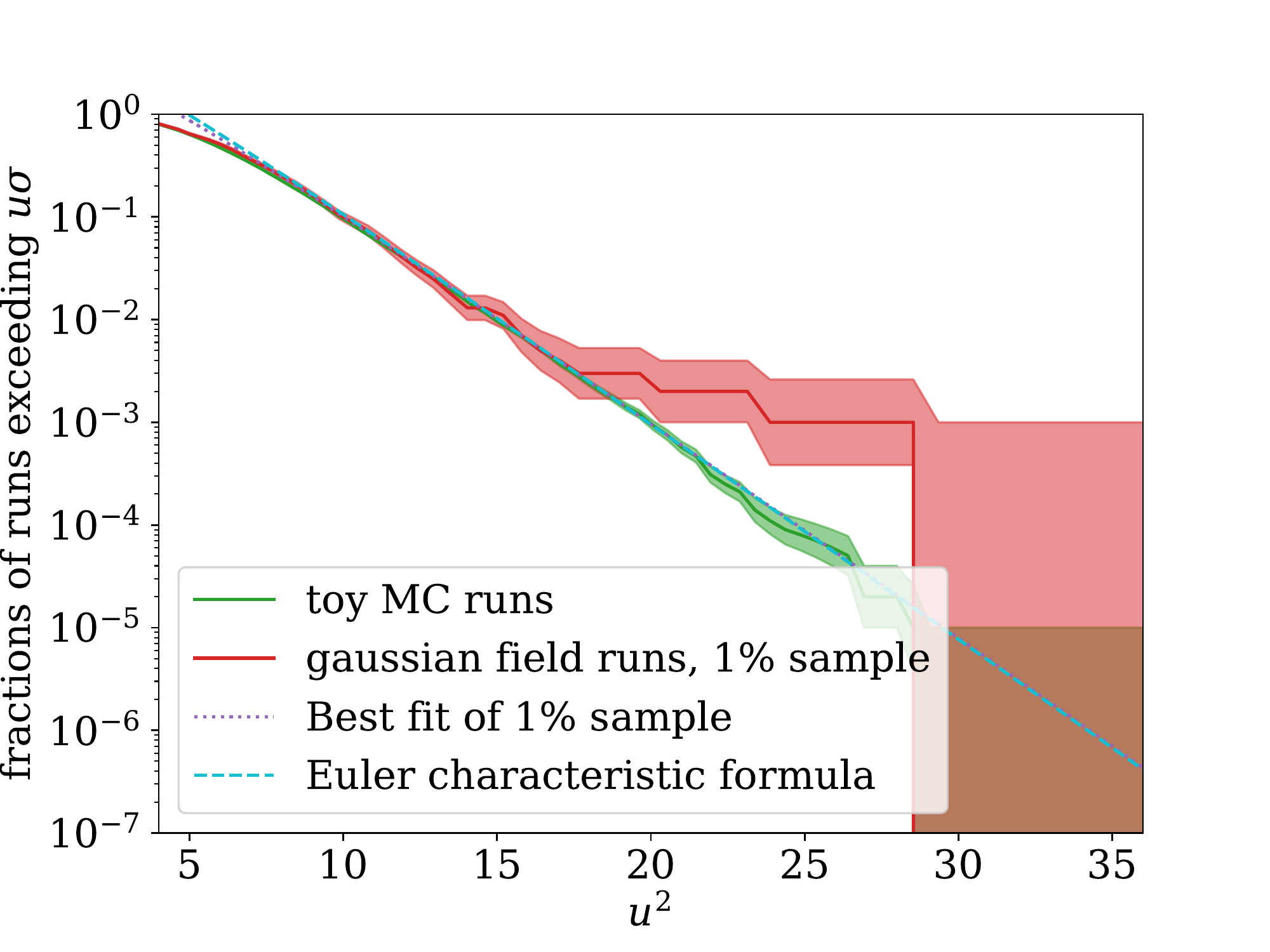}%
    \caption{The fraction of null random samples showing false positives as a function of the significance threshold in units of \(\sigma^2\). A fit using \autoref{eq:excursion_probability_approx} is shown (dotted purple), and we can see that a fit with only \(10^3\) samples agrees well with the excursion probability derived from \(10^5\) toy MC samples. We can see that the Euler characteristic computed using \autoref{eq:euler_characteristic} (dashed cyan) also agrees well with both the toy MC samples and the excursion probability fit.}\label{fig:2D_p_value_fit}
\end{figure}

As expected, the excursion probability obtained from toy Monte Carlo samples agree with those obtained by sampling Gaussian random fields. This demonstrates how Gaussian random fields can be sampled to produce large numbers of null significance map samples without a toy Monte Carlo whereby mock data is generated and used to produce significance maps via template matching. Finally, we can test the use of \autoref{eq:excursion_probability_approx} to fit the excursion probability. As the error term in \autoref{eq:excursion_probability_approx} is exponentially suppressed at small excursion probabilities, the fit only uses data points from after \(u^2 = 10\), where the excursion probability is approximately \(0.1\). An estimate using the covariance of the Gaussian kernel and \autoref{eq:euler_characteristic} is also performed; these are shown in \autoref{fig:2D_p_value_fit}. 

{\renewcommand{\arraystretch}{1.3}
\begin{table*}[htp]
    \centering
    \begin{tabular}{c|r@{\hspace{0em}}c@{\hspace{0em}}l|r@{\hspace{0em}}c@{\hspace{0em}}l}
         Method & \(p_{4\sigma}\) & & & \(p_{5\sigma}\) & &\\
         \hline
         Toy MC & $\left(6.27^{+0.25}_{-0.24}\right)$&$\times$& $10^{-3}$ & $\left(8.0^{+3.4}_{-2.4}\right)$&$\times$& $10^{-5}$ \\
         Gaussian random field & $\left(6.17^{+0.25}_{-0.24}\right)$&$\times$& $10^{-3}$ & $\left(6.0^{+3.0}_{-2.0}\right)$&$\times$& $10^{-5}$ \\
         Gaussian random field, spectral method & $\left(5.98^{+0.25}_{-0.24}\right)$&$\times$& $10^{-3}$ & $\left(5.0^{+2.8}_{-1.8}\right)$&$\times$& $10^{-5}$ \\
         Best fit & $6.36$&$\times$& $10^{-3}$ & $8.8$&$\times$& $10^{-5}$ \\
         Euler characteristic estimate & $6.40$&$\times$& $10^{-3}$ & $8.6$&$\times$& $10^{-5}$ \\
    \end{tabular}
    \caption{Global $p$-values at \(4\sigma\) and \(5\sigma\) local significance for the 2D toy problem. It can be seen that the $p$-values are consistent within stated binomial errors, and both the best fit value produced using a \(1\%\) sample size and the estimate produced with \autoref{eq:euler_characteristic} reproduce the simulated values well.}
    \label{tab:2D_gaussfield_metrics}
\end{table*}

Even though only \(10^3\) samples are used to fit the excursion probability, the fit matches the excursion probability expected from the toy MC samples. This demonstrates that fitting a limited set of samples using \autoref{eq:excursion_probability_approx} does indeed allow one to estimate the look-elsewhere effect correction with greatly reduced computational expense, in this case by a factor of $\sim 100$. These results are summarized in \autoref{tab:2D_gaussfield_metrics}, where it can be seen that the various methods all agree within expected uncertainties.

\section{Demonstration with a 1D template matching problem}\label{sec:1D_demo}

We showed in \autoref{sec:2D_demo} that in the case of a Gaussian search kernel and uniformly distributed underlying random variables, the covariance function can be easily computed. Indeed, while the derivation focused on Gaussian kernels, the result should hold in general for kernels that are closed under convolution, such as kernels that represent stable distributions \cite{Nolan_John_P_2020-09-13}.

In many cases, however, the covariance function might be easier to compute numerically. To demonstrate such an example, we consider a 1-dimensional search with a non-Gaussian kernel, for an excess in time-series data due to a particle interacting via a long-range force passing by an accelerometer. This toy problem is inspired by the Windchime project \cite{Windchime:2022whs}, where the direct detection of dark matter particles with masses of around the Planck mass will be attempted. While technically challenging, it has been suggested that this might be possible with large accelerometer arrays \cite{Carney:2019pza}. In the case of a dark matter particle passing by an accelerometer, the force as a function of time is given by \autoref{eq:force_accelerometer}, where \(G\) is the gravitational constant, \(m_\chi\) is the mass of a dark matter particle, \(m_s\) is the test mass of the sensor, \(b\) is the impact parameter of a dark matter track, and \(v\) is the velocity of the dark matter track \cite{Carney:2019pza}.
\begin{equation}\label{eq:force_accelerometer}
    F(t) = \frac{G m_{\chi} m_s b}{\left(b^2 + v^2 t^2\right)^{3/2}}
\end{equation}
For template matching purposes, a normalized template with the same shape as \autoref{eq:force_accelerometer} can be used. This is shown in equation \autoref{eq:force_template}.
\begin{equation} \label{eq:force_template}
    f(t) = \frac{v b^2}{2\left(b^2 + v^2 t^2\right)^{3/2}}
\end{equation}

\begin{figure}
 \centering
    \includegraphics[width=\columnwidth]{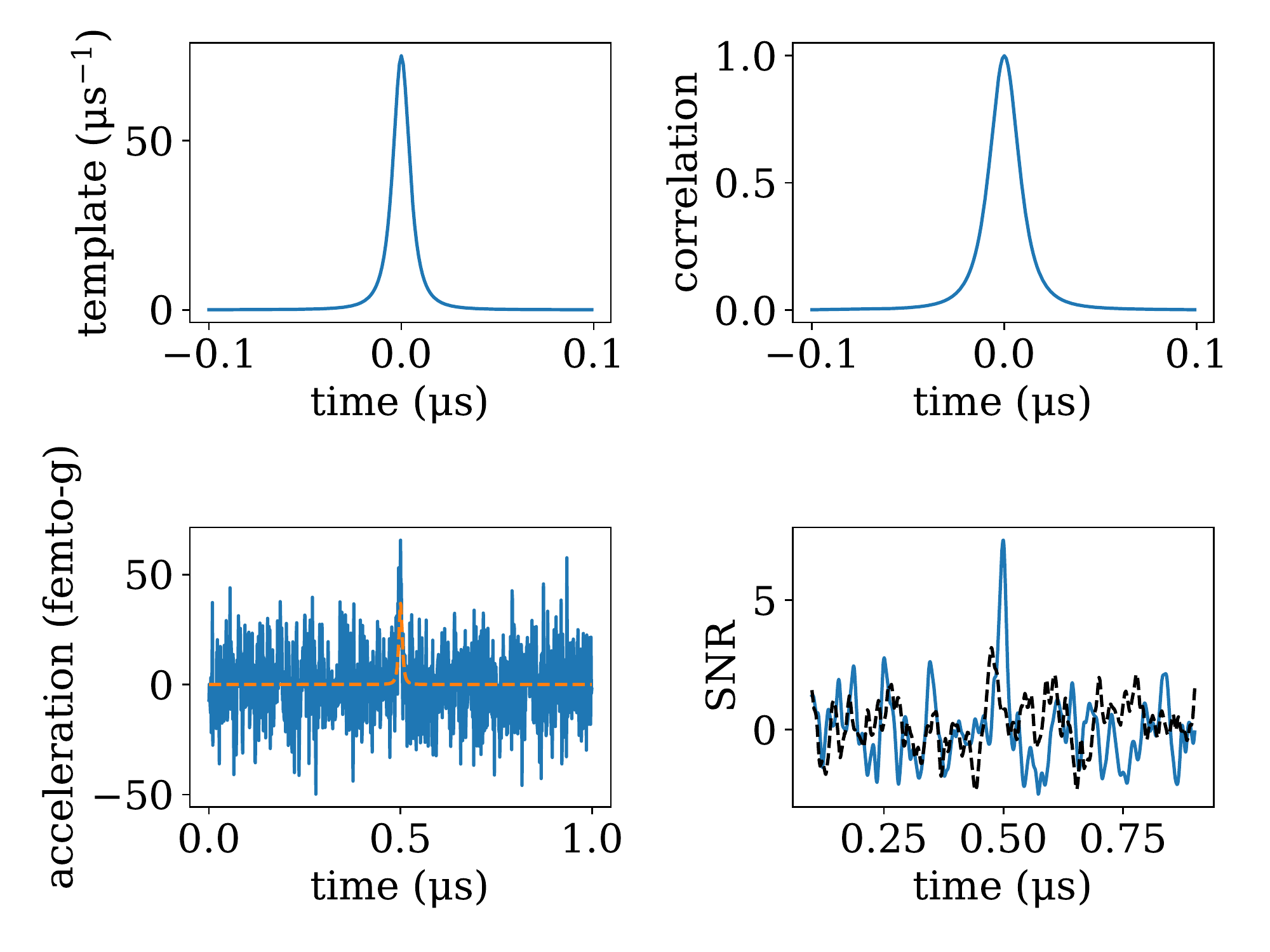}%
    \caption{Top left: The template function of the 1D toy problem. Top right: The correlation function derived as the autocorrelation of the template. Bottom left: One random sample containing a true signal, with the signal truth expectation shown in dashed orange. Bottom right: Two significance maps generated using the toy MC procedure. The blue line contains a true signal, whereas the black dashed line does not.}\label{fig:1D_4_plots}
\end{figure}

In this demonstration, values of \(b=3 \unit{mm}\) and \(v=3\times10^5 \unit{m/s}\) are used, and the sampling rate is \(10^9 \unit{Hz}\). With this information, we can generate the template for template matching, and then compute the covariance function using \autoref{eq:covariance_two_point_indep}. As this system is also described by a stationary random field, this is done by computing the autocorrelation of the template. These are shown together with MC data samples in \autoref{fig:1D_4_plots}. 


\begin{table*}[htp]
    \centering
    \begin{tabular}{c|r@{\hspace{0em}}c@{\hspace{0em}}l|r@{\hspace{0em}}c@{\hspace{0em}}l}
         Method & \(p_{4\sigma}\) & & & \(p_{5\sigma}\) & &\\
         \hline
         Toy MC & $\left(5.82^{+0.25}_{-0.24}\right)$ &$\times$& $10^{-3}$ & $\left(3.0^{+2.3}_{-1.3}\right)$&$\times$& $10^{-5}$ \\
         Gaussian random field & $\left(6.13^{+0.25}_{-0.24}\right)$&$\times$& $10^{-3}$ & $\left(10^{+4}_{-3}\right)$&$\times$& $10^{-5}$ \\
         Gaussian random field, spectral method & $\left(6.29^{+0.25}_{-0.25}\right)$&$\times$& $10^{-3}$ & $\left(10^{+4}_{-3}\right)$&$\times$& $10^{-5}$ \\
         Best fit & $5.42$&$\times$& $10^{-3}$ & $4.9$&$\times$& $10^{-5}$ \\
         Euler characteristic estimate & $7.56$&$\times$& $10^{-3}$ & $8.4$&$\times$& $10^{-5}$ \\
    \end{tabular}
    \caption{Global $p$-values at \(4\sigma\) and \(5\sigma\) local significance for the 1D template matching problem. It can be seen that the $p$-values are consistent within stated binomial errors, and both the best fit value produced using a \(1\%\) sample size and the estimate produced with \autoref{eq:euler_characteristic} reproduce the simulated values well.}
    \label{tab:1D_gaussfield_metrics}
\end{table*}

As in \autoref{sec:2D_demo}, the global $p$-values at \(4\sigma\) and \(5\sigma\) local significance are computed using \(10^5\) signal-free samples each using a toy MC, direct sampling of the Gaussian random field using the covariance function, sampling of the Gaussian random field in frequency space, and a best-fit with \(10^3\) samples using \autoref{eq:excursion_probability_approx}. The best-fit only uses data points after \(u^2=10\), where the excursion probability is approximately \(0.1\). These results are shown in \autoref{tab:1D_gaussfield_metrics}. 

As we expect, the different values agree with the computed uncertainty, demonstrating how the methods outlined in this paper can be used to estimate the look-elsewhere effect. While the example here uses a template that is relevant to the Windchime project, this procedure can be used in general to calibrate the look-elsewhere effect correction for problems involving template matching or matched filtering of time-series data, including sonar \cite{Abraham2019} and fast radio burst detection \cite{Petroff:2021wug}. It should be noted that for cases involving multiple templates, correlations between templates would need to be computed as well to avoid underestimation of the significance of a signal.

\section{The look-elsewhere effect in a peak search with profile likelihoods}\label{sec:peak_search}

Profile likelihoods are often used in particle physics experiments as part of likelihood ratio tests~\cite{Algeri:2019lah}. In this section, we explore the use of Gaussian random fields to model the look-elsewhere effect correction for a peak search using a likelihood ratio test. This type of statistical problem can be encountered in dark matter searches such as~\cite{XENON:2020rca, XENON:2022ltv}, or resonance searches in collider experiments~\cite{ATLAS:2016gzy, CMS:2016kgr}.

To demonstrate how Gaussian random fields can be used to model the behaviour of the likelihood ratio test statistic, we consider a peak search in a dataset with no signal and with a background distribution that is flat, aside from an efficiency roll-off at the start. This is modelled using the cumulative distribution function of a gamma distribution. The level of the likelihood function is left as a nuisance variable for the profile likelihood.

\begin{figure}
 \centering
    \includegraphics[width=\columnwidth]{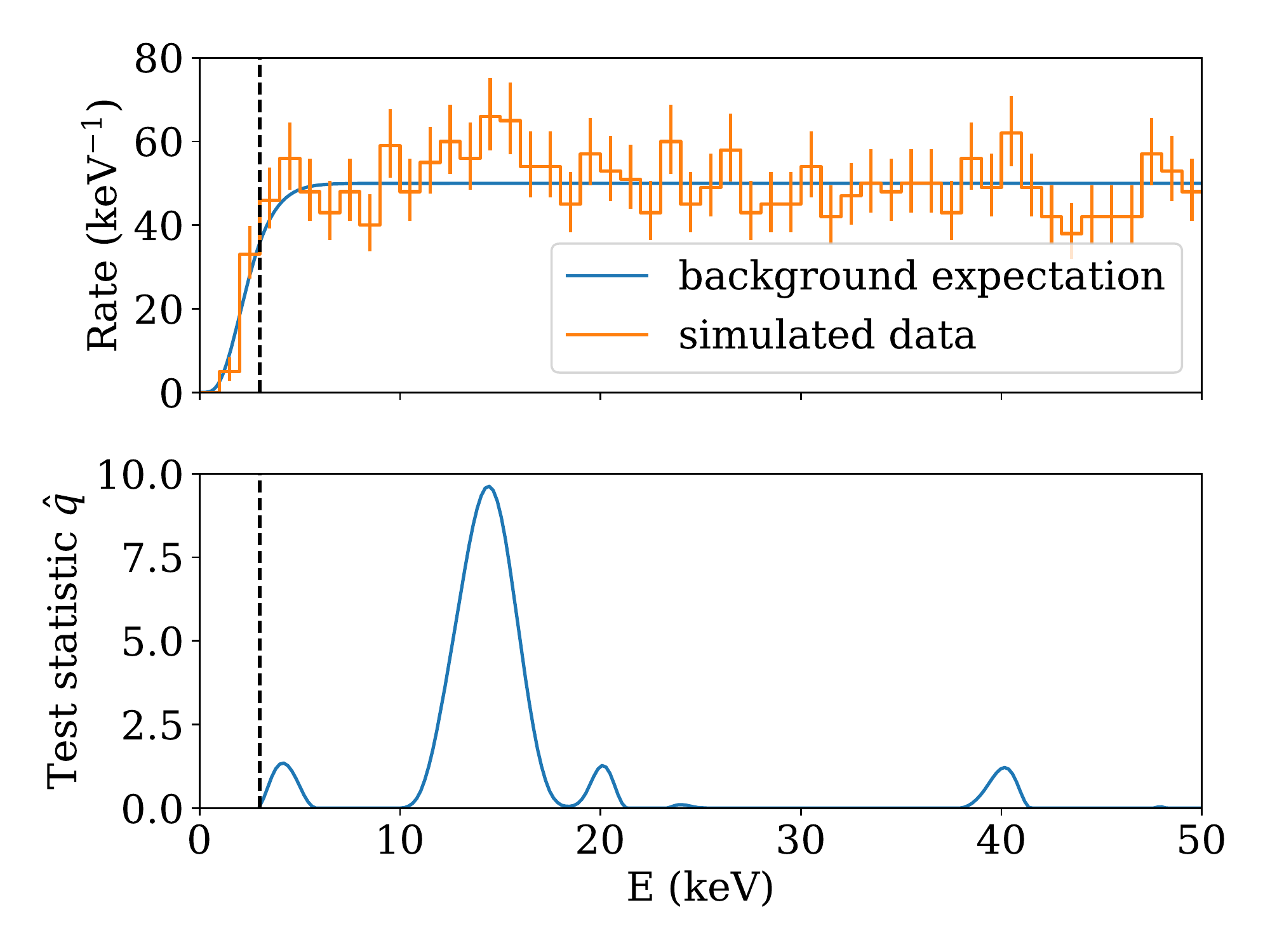}%
    \caption{Top: One data sample from the MC simulation of a peak search and the background expectation. $1\sigma$ errorbars are shown. Bottom: The test statistic computed using \autoref{eq:profile_lh_test_statistic}. The lower energy threshold for the peak search is indicated by the black dashed line. The width of the Gaussian peak is \(1 \unit{keV}\) for this search.}\label{fig:resonance_search_toymc}
\end{figure}

An unbinned likelihood is used, given by~\cite{Algeri:2019lah}:
\begin{equation}
\begin{split}
    \mathcal{L}(E_{peak}, \mu, \beta) =& \frac{1}{\mu + \beta} \mathrm{Poisson}(N|\mu + \beta) \times \\
    &\prod^N_{i=1}\left(\beta B(E_i) + \mu \mathrm{Gauss}(E_i - E_{peak})\right),
\end{split}
\end{equation}
where $\mu$ is the signal level, $\beta$ is the background level, $E_{peak}$ is the signal location, $N$ is the number of events in the dataset, $B(E)$ is the background distribution, and $\mathrm{Gauss}(E)$ represents the standard Gaussian distribution. The profile likelihood is then given by $\mathcal{L}(E_{peak}, \mu, \hat{\beta})$, where $\hat{\beta}$ is the best-fit value of $\beta$ given $(E_{peak}, \mu)$. The test-statistic is thus:
\begin{equation}\label{eq:profile_lh_test_statistic}
    \hat{q}(E_{peak}) = -2\frac{\mathcal{L}(E_{peak}, 0, \hat{\beta}_0)}{\mathcal{L}(E_{peak}, \hat{\mu}, \hat{\beta})}.
\end{equation}
A plot of $\hat{q}(E_{peak})$ can be seen in \autoref{fig:resonance_search_toymc}.

\begin{figure}
 \centering
    \includegraphics[width=\columnwidth]{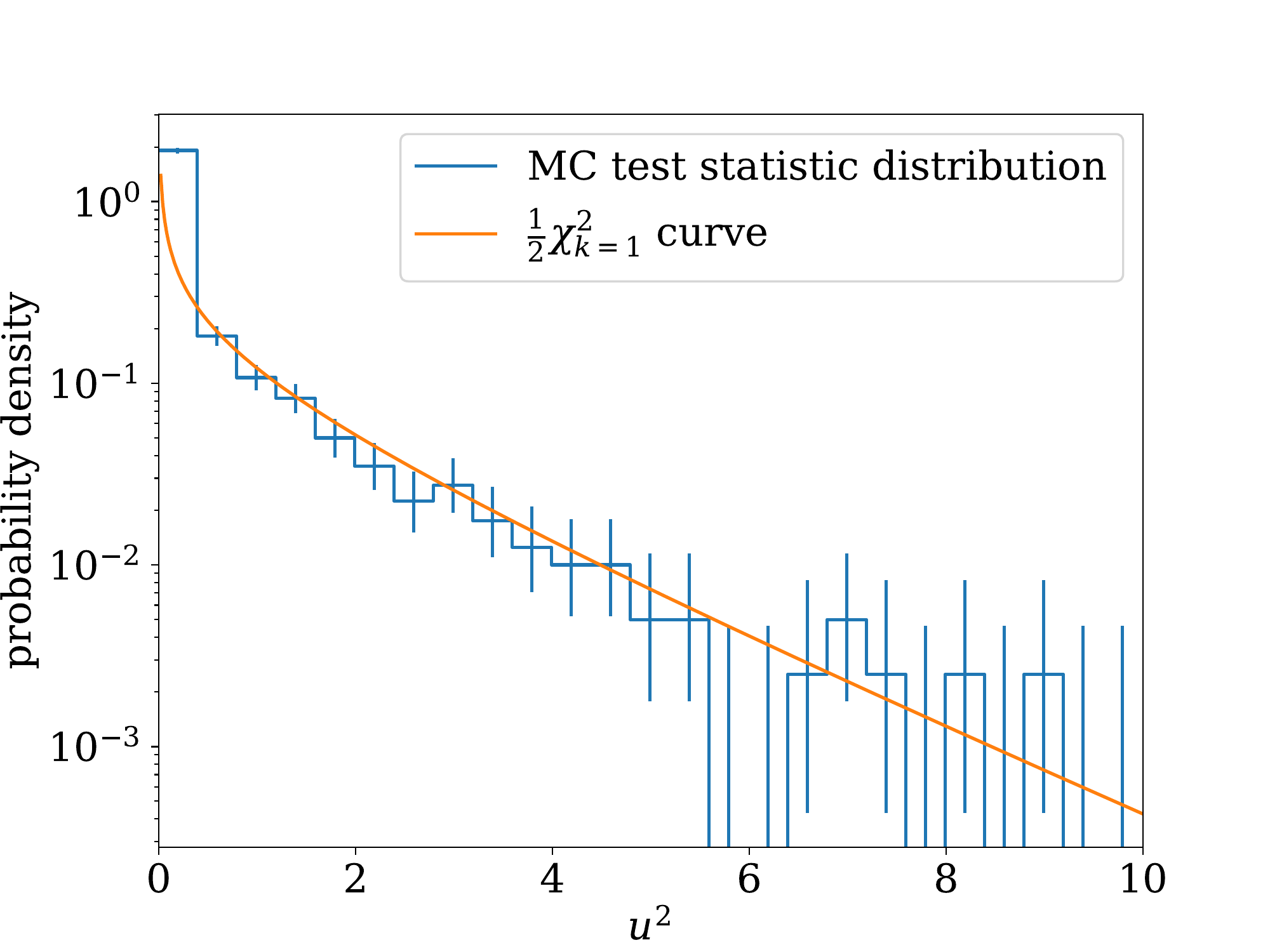}%
    \caption{The distribution of the test statistic computed with \autoref{eq:profile_lh_test_statistic} at random energies for $1000$ simulated data samples. It can be seen that aside from the first bin, the distribution follows the $\frac{1}{2}\chi^2_{k=1}$ curve.}\label{fig:resonance_search_asymptoticity}
\end{figure}

According to Wilks' theorem, the likelihood ratio test statistic should asymptotically approach a $\chi^2$ distribution. However, as the null-hypothesis is defined by $\mu_0 = 0$, and this lies on the edge of the parameter space $\mu \in [0,\infty)$, the asymptotic distribution is instead $\frac{1}{2}\chi^2_{k=1} + \frac{1}{2}\delta(0)$~\cite{Algeri:2019lah}. The probability of an upward fluctuation of this this distribution above $u^2$ is equal to the probability of an upward fluctuation of a standard normal distribution above $u$. Before we can model the test statistic using Gaussian random fields, however, asymptoticity must first be verified. This can be done by sampling the test statistic for different toy MC datasets at random values of $E_{peak}$, and verifying that the resulting distribution follows the expected distribution. This is shown in~\autoref{fig:resonance_search_asymptoticity}.

\begin{figure}[htp]
 \centering
    \includegraphics[width=\columnwidth]{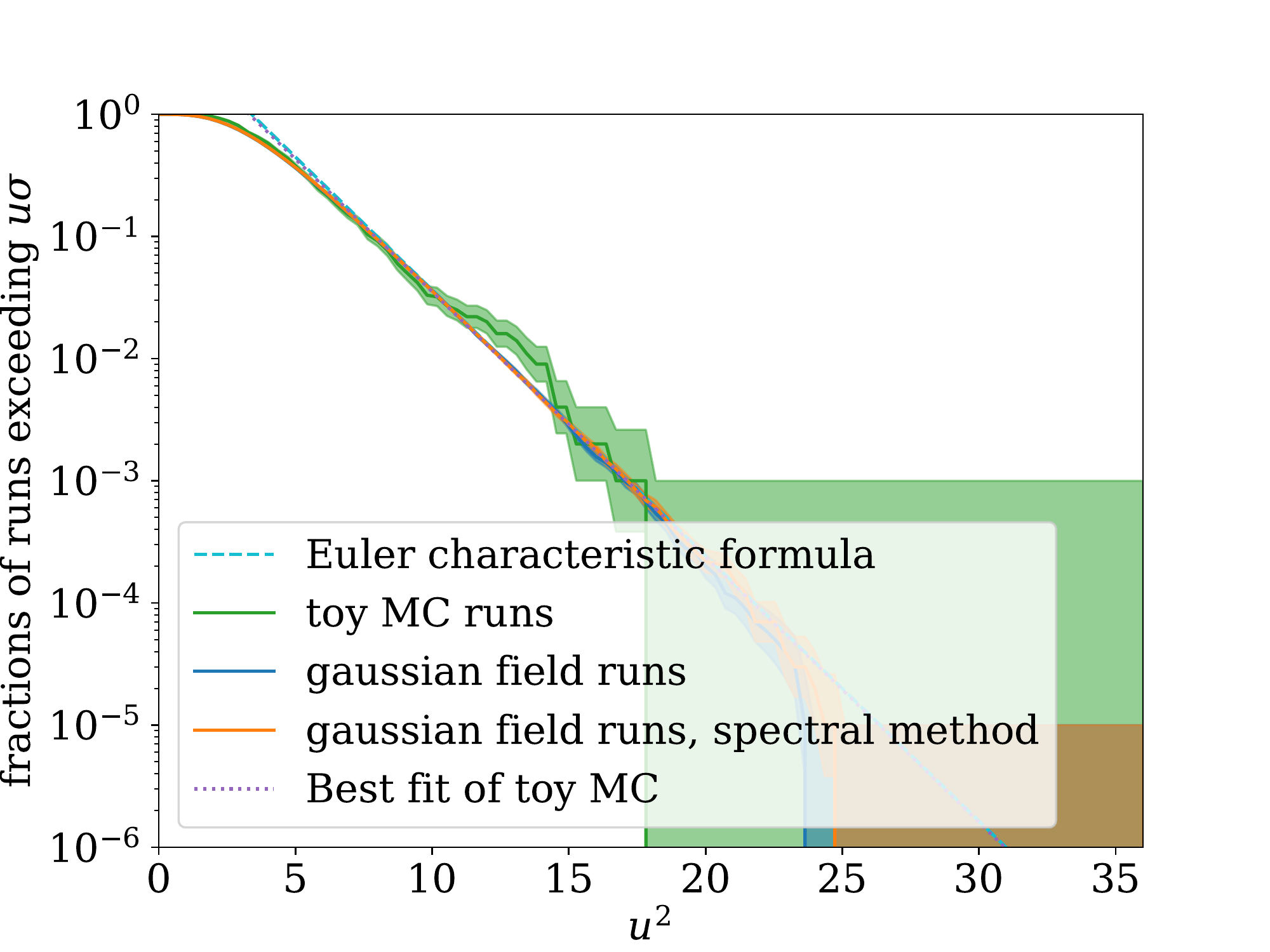}%
    \caption{The fraction of null random samples showing false positives as a function of the significance threshold in units of \(\sigma^2\). A fit using \autoref{eq:excursion_probability_approx} is conducted on the toy MC samples. The toy MC was only run $10^3$ times, due to the computational expense. We can see that the Euler characteristic computed using \autoref{eq:euler_characteristic} (dashed cyan) agrees well with both the toy MC samples and the excursion probability fit. In addition, the excursion probability obtained from sampling the Gaussian random field $10^5$ times both naively and using the spectral method agree as well.}\label{fig:resonance_search_p_value_fit}
\end{figure}

\begin{table*}[htp]
    \centering
    \begin{tabular}{c|r@{\hspace{0em}}c@{\hspace{0em}}l|r@{\hspace{0em}}c@{\hspace{0em}}l}
         Method & \(p_{4\sigma}\) & & & \(p_{4.5\sigma}\) & &\\
         \hline
         Toy MC & $\left(2^{+2}_{-1}\right)$ &$\times$& $10^{-3}$ & $\left(0^{+10}_{-0}\right)$&$\times$& $10^{-4}$ \\
         Gaussian random field & $\left(1.57^{+0.13}_{-0.12}\right)$&$\times$& $10^{-3}$ & $\left(1.7^{+0.5}_{-0.4}\right)$&$\times$& $10^{-4}$ \\
         Gaussian random field, spectral method & $\left(1.83^{+0.14}_{-0.13}\right)$&$\times$& $10^{-3}$ & $\left(2.2^{+0.5}_{-0.4}\right)$&$\times$& $10^{-4}$ \\
         Best fit & $1.8$&$\times$& $10^{-3}$ & $2.1$&$\times$& $10^{-4}$ \\
         Euler characteristic estimate & $1.8$&$\times$& $10^{-3}$ & $2.2$&$\times$& $10^{-4}$ \\
    \end{tabular}
    \caption{Global $p$-values at \(4\sigma\) and \(4.5\sigma\) local significance for the peak search with profile likelihoods. It can be seen that the $p$-values are consistent within stated binomial errors, and both the best fit value produced using a fit of the toy MC samples and the estimate produced with \autoref{eq:euler_characteristic} match the sampled values well. Local significances of \(4\sigma\) and \(4.5\sigma\) are shown here instead of \(4\sigma\) and \(5\sigma\) for the earlier examples as the look-elsewhere effect correction is smaller here, resulting in very large errorbars at \(5\sigma\) local significance.}
    \label{tab:resonance_gaussfield_metrics}
\end{table*}

We can now compare the excursion probability from toy MC runs, such as that shown in \autoref{fig:resonance_search_toymc}, with the excursion probability obtained using Gaussian random fields. This is shown in \autoref{fig:resonance_search_p_value_fit}. For this comparison, only $1000$ runs of the toy MC were done due to the computational expense. As the signal model is a Gaussian peak with \(\sigma=1\unit{keV}\), the covariance function is also modelled as a Gaussian peak with \(\sigma=\sqrt{2} \unit{keV}\), as described in \autoref{sec:2D_demo}. The results are summarised in \autoref{tab:resonance_gaussfield_metrics}. The best-fit only uses data points after \(u^2=8\), where the excursion probability is approximately \(0.1\). It can be seen that as with the earlier examples, the various methods agree within uncertainties. As such, it can be seen that these methods can be used to determine the look-elsewhere effect correction for a likelihood ratio test using profile likelihoods, as long as one verifies that the test statistic follows the asymptotic distribution.

\section{Application to Windchime}\label{sec:windchime_demo}

Finally, these methods can now be applied to estimate the look-elsewhere effect correction needed for a dark matter direct detection experiment based on the Windchime concept \cite{Windchime:2022whs}. In this section, we consider the detection of dark matter interacting via a long range force using a \(0.6 \unit{m}\) array of \(4^3\) accelerometers, with a sampling rate of \(10^7 \unit{Hz}\).

The force on a single sensor is given in \autoref{eq:force_accelerometer}. However, for a particle passing through a sensor array, the impact parameter \(b\) would be different for each sensor, and additionally, the time of closest approach differs between sensors. Thus, instead of using the template for a single sensor, a template for the entire array is considered. As each template represents a track, we have to consider the parameterization of a track through the sensor array. We accomplish this using a bounding sphere that is larger than the accelerometer array, so that each track through the array intersects the bounding sphere twice and hence can be parameterized by two points on the bounding sphere. Any given template can then be parameterized by 6 parameters: velocity \((v)\), entry time, the spherical coordinates of the entry point \((\cos(\theta_0), \phi_0)\), and the spherical coordinates of the exit point \((\cos(\theta_1), \phi_1)\). The cosine of the \(\theta\) angles is used as evenly-spaced bins in \(\cos(\theta)\) represent equal-sized areas on a sphere. Here, we consider a bounding sphere with a diameter of \(1 \unit{m}\), so that it encloses the entire array.

\begin{figure}
 \centering
    \includegraphics[width=\columnwidth]{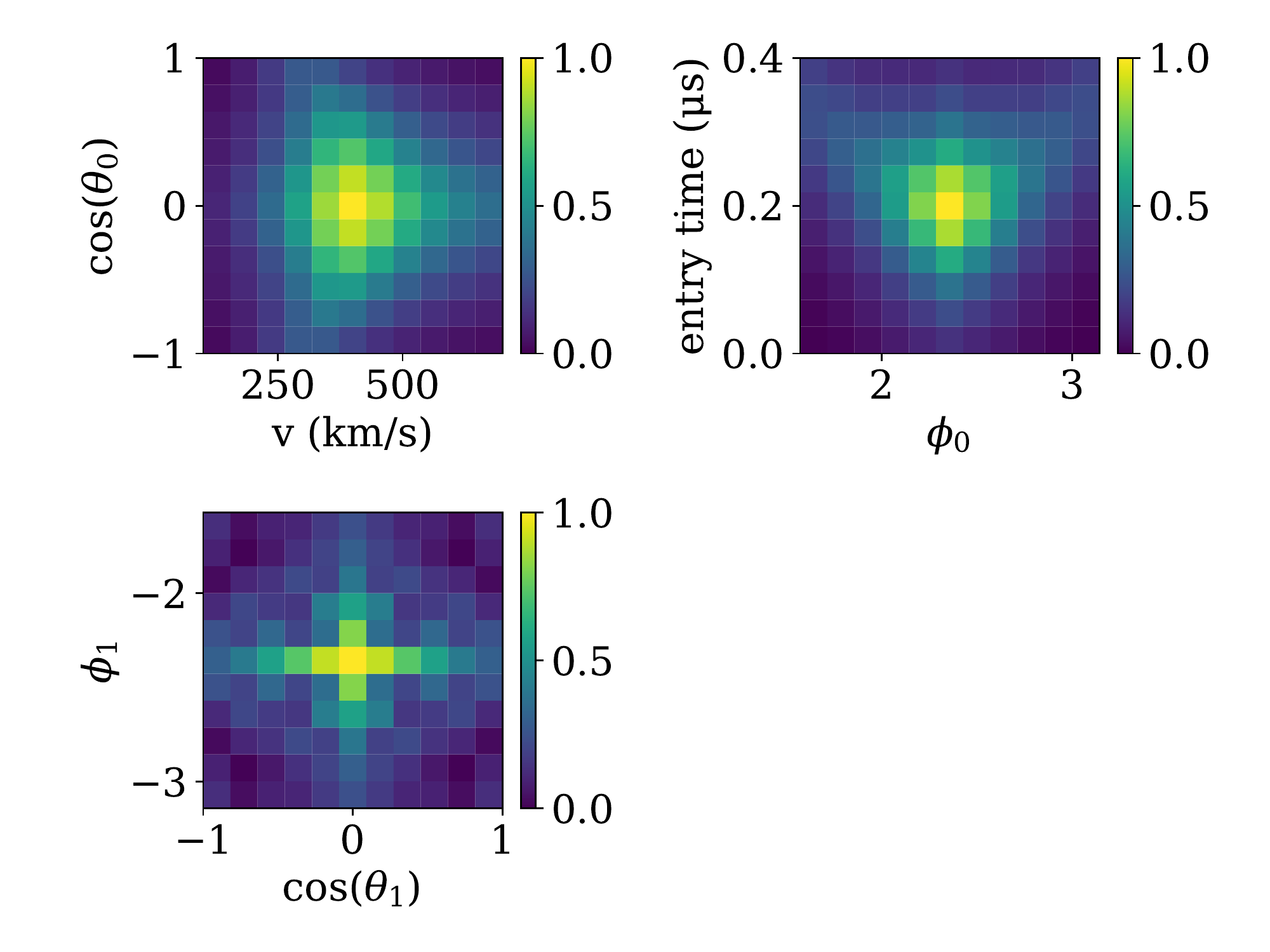}%
    \caption{2D slices of the 6D covariance function computed using accelerometer array templates.}\label{fig:windchime_correlation_slices}
\end{figure}

For each set of the 6 parameters, a template is generated by considering the force on each sensor over a series of timesteps. At every timestep, the distance between the particle and every sensor is calculated, and the template for each sensor is computed using the inverse-square law. The equation for the template at the \(i^\mathrm{th}\) timestep and the \(j^\mathrm{th}\) sensor is thus:
\begin{equation} \label{eq:array_template}
    \mathbf{f}_{ij} = \frac{\mathbf{r}_{ij}}{r_{ij}^3}
\end{equation}
After the computation of the entire template over a set of timesteps and all sensors, the template is divided by its sum to normalize it to unity. Finally, the covariance between two sets of parameters can be computed by summing across all sensors and timesteps using \autoref{eq:covariance_two_point_indep}. This allows for the covariance function to be mapped out between one chosen template and other templates in the parameter space. Some 2D slices of the covariance function are shown in \autoref{fig:windchime_correlation_slices}.

\begin{figure}
 \centering
    \includegraphics[width=\columnwidth]{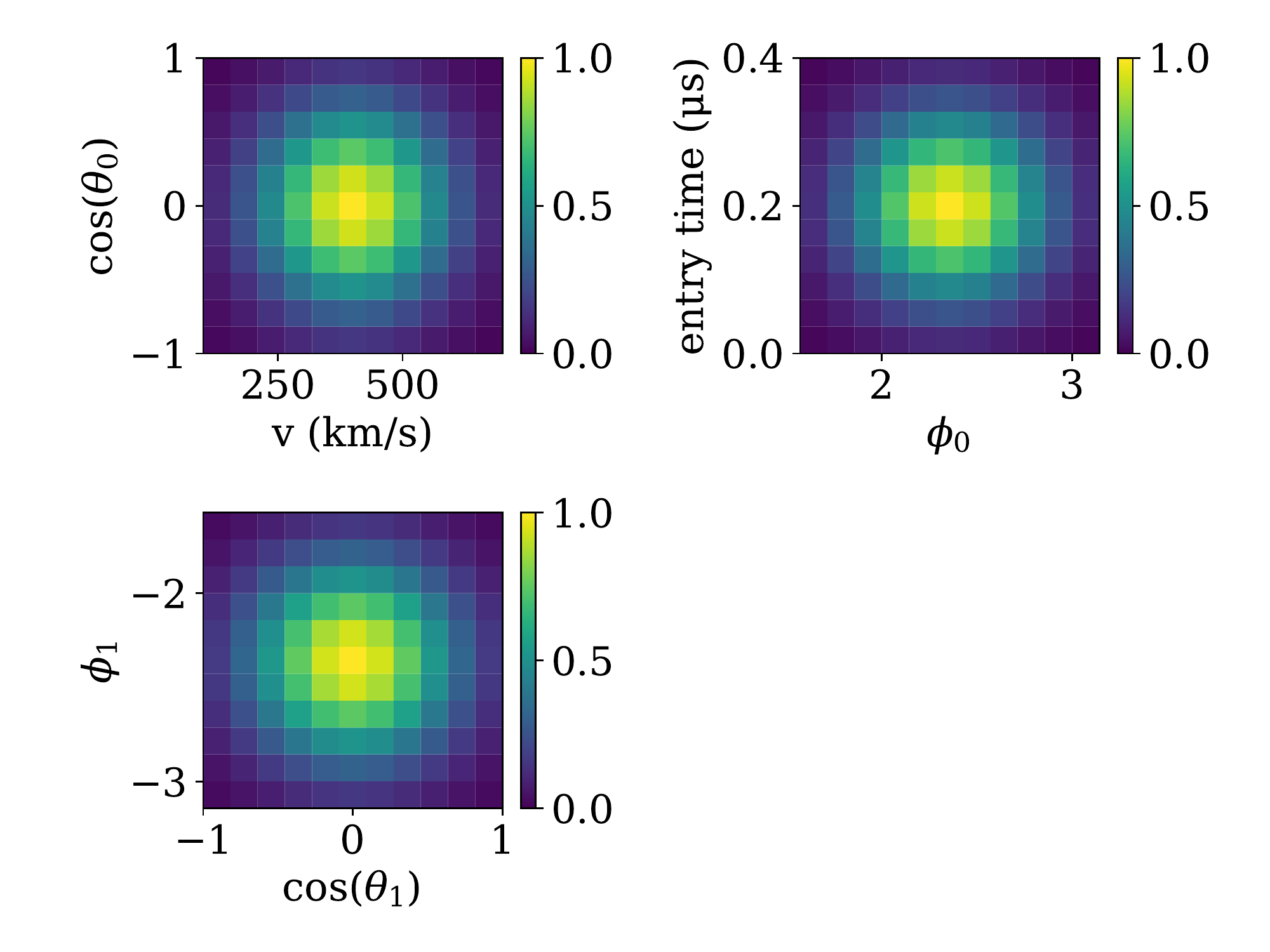}%
    \caption{Gaussian kernel approximation of the slices of the 6D correlation function shown in Fig.~\ref{fig:windchime_correlation_slices}}\label{fig:windchime_correlation_slices_approximate}
\end{figure}

It can be seen from \autoref{fig:windchime_correlation_slices} that the Gaussian random field representing this problem is not stationary. This is because in the case of a stationary field, the covariance function only depends on the displacement between points, as described in \autoref{eq:stationary_covariance}. This implies that \(K(\mathbf{x}, \mathbf{x}-\mathbf{s}) = K(\mathbf{x}-\mathbf{s}, \mathbf{x}) = K(\mathbf{x}, \mathbf{x} + \mathbf{s})\). Thus, for a stationary process, \(K_s(\mathbf{s}) = K_s(-\mathbf{s})\) and the covariance function is symmetric. Unfortunately, this means that Wiener-Khinchin theorem~\cite{Vetterli:2014-10-20, Jaynes_E_T_2003-04-10} does not apply, and spectral sampling of this covariance is not possible. To get an estimate of the look-elsewhere correction, we can approximate the covariance function using a symmetric functional form. Here, for the purposes of an order-of-magnitude estimate, we use a Gaussian kernel to approximate the covariance function. To do this, first, points are sampled based on the 6D histogram shown in~\ref{fig:windchime_correlation_slices}. After that, the covariance of this large point cloud is computed to find the approximate covariance function. The 2D slices of this approximate covariance function are shown in \autoref{fig:windchime_correlation_slices_approximate}.

\begin{figure}
 \centering
    \includegraphics[width=\columnwidth]{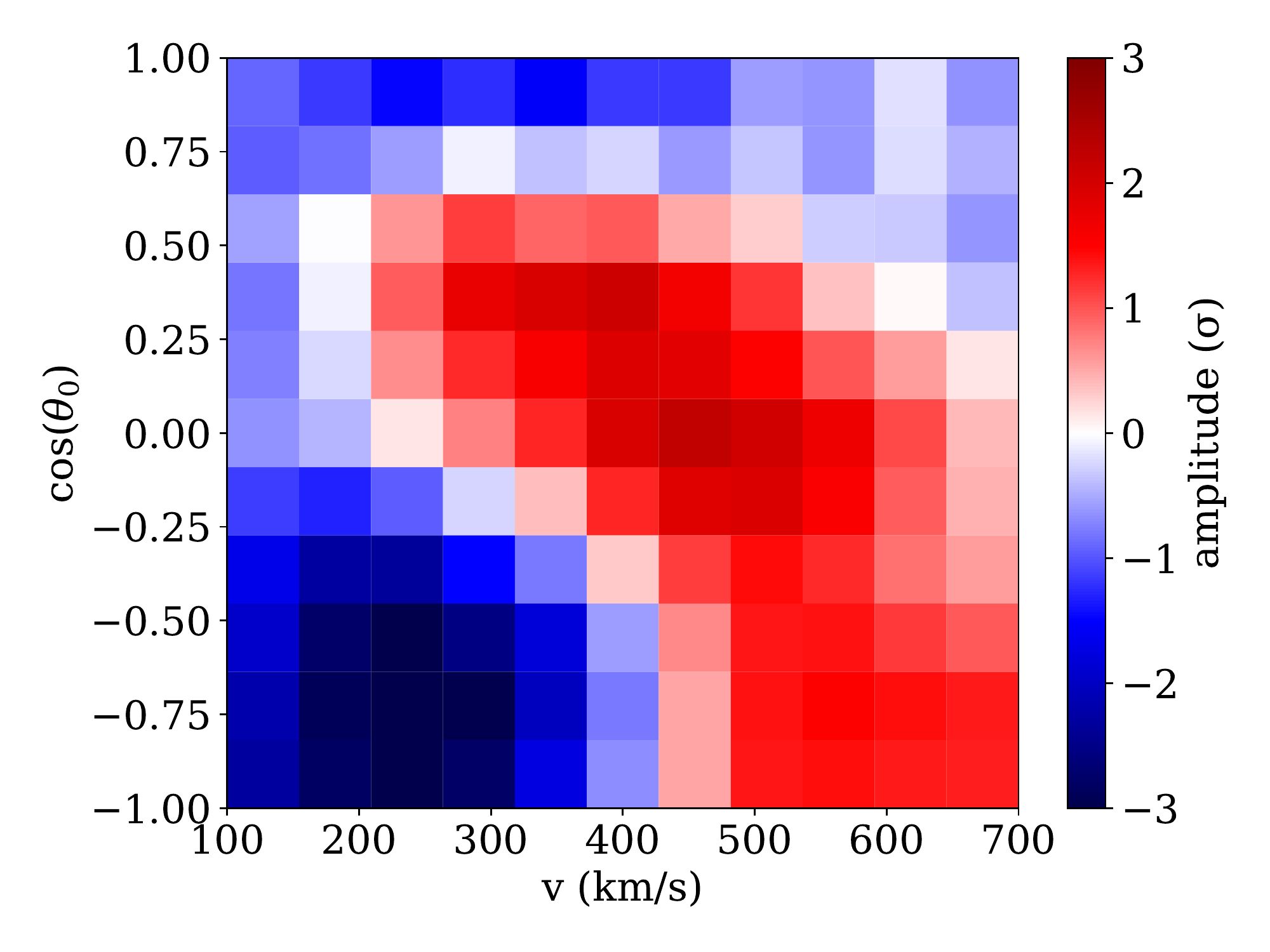}%
    \caption{2D slice of one null sample, generated using the covariance function shown in Fig.~\ref{fig:windchime_correlation_slices_approximate}.}\label{fig:windchime_sample}
\end{figure}

A random sample from the Gaussian random field represented by \autoref{fig:windchime_correlation_slices_approximate}, sampled using the spectral method, is shown in \autoref{fig:windchime_sample}. Random samples are generated approximating the parameter space with a Euclidean parameter space of equal volume. This is conservative, as correlations near the edges of the parameter space that should be connected, such as $\phi_0=-\pi$ and $\phi_0=\pi$, that are not considered would result in a lower trial factor if this approximation was not taken. Thus, the trial factor inferred with this approximation is higher than it would otherwise be.

\begin{figure}
 \centering
    \includegraphics[width=\columnwidth]{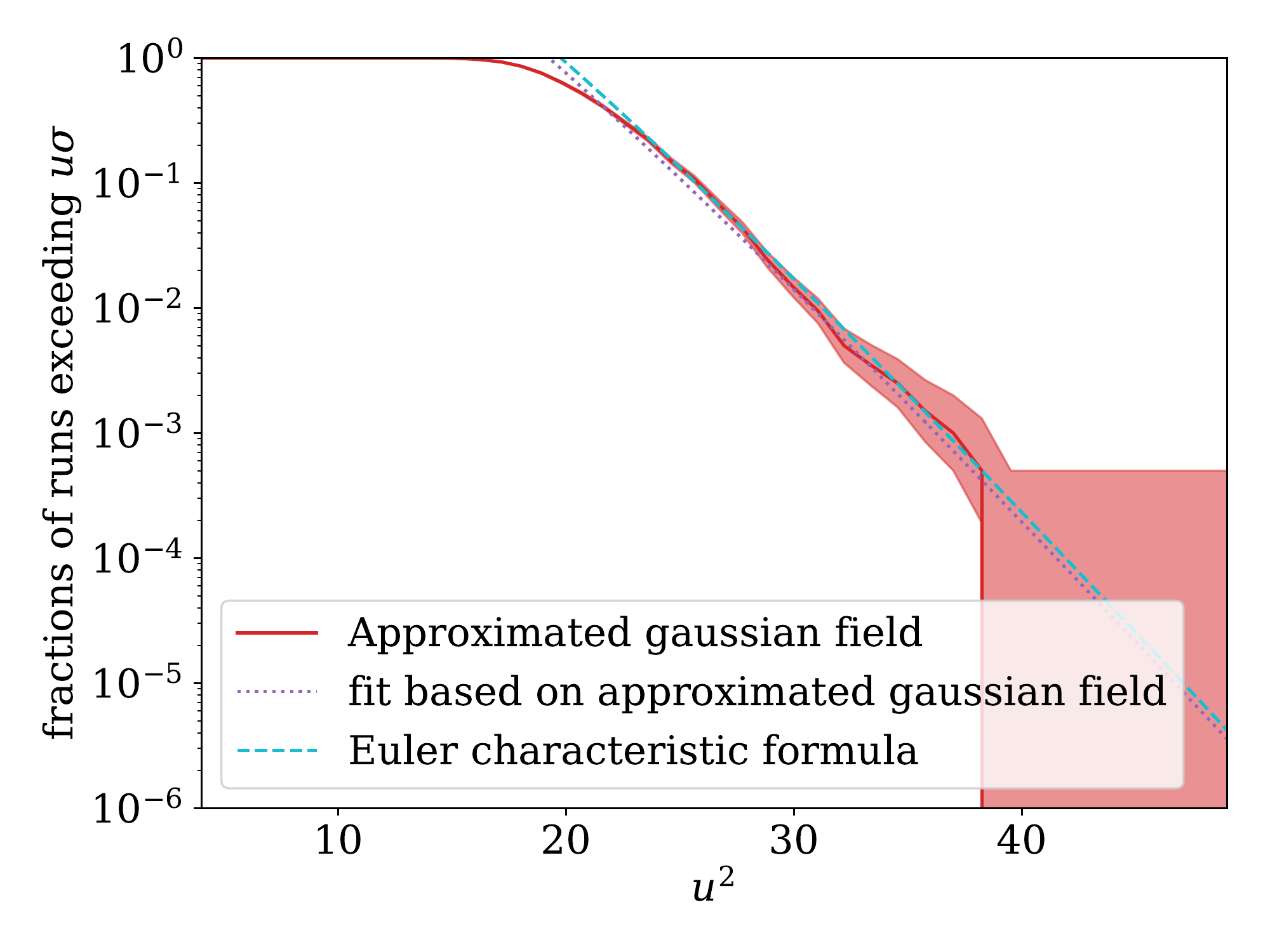}%
    \caption{The fraction of null random samples showing false positives as a function of the significance threshold in units of \(\sigma^2\), with a fit using \autoref{eq:excursion_probability_approx} (dotted purple) and the Euler characteristic (\autoref{eq:euler_characteristic}) (dashed cyan). \(2000\) samples generated using the spectral method are shown here.}\label{fig:windchime_p_value_fit}
\end{figure}

The excursion probability can now be fitted to samples such as \autoref{fig:windchime_sample}. The excursion probability estimated with \(2000\) such samples is shown in \autoref{fig:windchime_p_value_fit}.

We can now compute the trial factor using the fit in \autoref{fig:windchime_p_value_fit}. First, we need to compute the signal-to-noise ratio threshold needed for a search with confidence level \(1-\alpha\) over time \(t\). Here, we use \(\alpha=0.0027\), corresponding to a significance level of \(3\sigma\). The signal-to-noise ratio threshold is then found by solving \autoref{eq:SNR_threshold} for \(u\):
\begin{equation}\label{eq:SNR_threshold}
    \Psi(u)\frac{V'}{V}\frac{T'}{T} - \alpha = 0
\end{equation}
where \(\Psi(u)\) is the fitted excursion probability function, \(\frac{V'}{V}\) corresponds to the fraction of parameter space covered by the sampled Gaussian field, and \(\frac{T'}{T}\) corresponds to the search time covered by the random field divided by the desired search time. This procedure tells us that we need a signal-to-noise ratio threshold of at least \(8.4\) for a \(1 \unit{s}\) search time and \(10.4\) for a \(1\unit{yr}\) search time. The trial factor, \(N_\text{trials}\), is given by
\begin{equation}\label{eq:N_trials}
    N_{trials} = \frac{\alpha}{\Phi(u)}.
\end{equation}
This results in estimated trial factors of \(\sim 10^{14}\) for a \(1 \unit{s}\) search, and \(\sim 10^{22}\) for a \(1\unit{yr}\) search. We can see that due to the high dimensional search space, the Windchime experiment suffers from a rather high trial factor. Thus, thresholds much higher than the \(5\sigma\) level customary in particle physics \cite{ParticleDataGroup:2022pth, Lyons:2013yja, Barlow:2019svl} are needed for a rare event search with an accelerometer array.

\subsection{The computational expense of calibrating for the look-elsewhere effect}

For a $N$ dimensional parameter space with $m$ bins per dimension, computing the covariance function requires comparing one parameter with $N^m-1$ others. Each comparison is equivalent to a template matching computation or likelihood function evaluation in terms of computational effort. This is a significant speedup over sampling the random field directly via toy MC, as each individual toy MC sample would involve generating a mock dataset and then doing template matching across all sensors and timesteps for $N^m$ different sets of parameters. In the case shown here with 6 parameters and 11 bins per dimension, computing the convariance function required $1.7\times10^6$ such evaluations.

This also compares favourably with frequentist-Bayesian methods for the estimation of the look-elsewhere effect such as~\cite{Bayer:2020pva}. This is because such methods still require a posterior distribution to be sampled, which still requires a large number of likelihood evaluations. While a detailed analysis of the best-case number of likelihood evaluations required is beyond the scope of this paper, we found that searching for a track in simulated data for the sensor array described in this section required approximately $\sim10^7$ likelihood evaluations using nested sampling with a slice sampler as implemented in~\cite{Buchner2021UltraNestA}. However, it should be noted that if the posterior was to be sampled anyway as part of the data analysis and inference procedure, the method presented in~\cite{Bayer:2020pva} does allow one to calibrate for the look-elsewhere effect with minimal computational overhead.

We can also compute the look-elsewhere effect correction with minimal computational expense if a search for signals is conducted using the self-calibration method presented in~\cite{Bayer:2021lhk}. However, the methods shown in this work allow for us to directly compute this correction without having to perform such a search. In the cases shown in \autoref{sec:2D_demo}, \autoref{sec:1D_demo}, and \autoref{sec:peak_search}, the look-elsewhere effect correction can be directly computed from~\autoref{eq:euler_characteristic} using the search template or signal model, and in this section, while the covariance function had to be computed numerically, this only required sampling a small subset of the parameter space. As such, while the self-calibration method shown in~\cite{Bayer:2021lhk} is extremely performant and allows one to compute the look-elsewhere effect correction with almost no computational overhead when such a search is conducted, the methods presented in this work are more useful for quickly estimating the look-elsewhere effect for the purposes of a sensitivity study where such a search has not been conducted, and may be prohibitive without additional computational infrastructure.

\section{Conclusions}

In this paper, we described and demonstrated the use of Gaussian random fields in the estimation of the look-elsewhere effect. The presented methods can be used to greatly reduce the computational requirements for the estimation of the look-elsewhere effect. This is particularly useful for high-dimensional and otherwise computationally complex problems. Our methods can also be helpful for sensitivity projections of future experiments, where the computational infrastructure needed for the data-analysis of such an experiment does not yet exist.

We have shown that Gaussian random fields can be used to model a large set of statistical problems commonly encountered in physics.
When it has been ascertained that a given significance map can be modelled by a Gaussian random field, three techniques that can be used to reduce the computational cost of estimating the look-elsewhere effect correction for local significance are demonstrated in this paper. First, various methods exist for the efficient sampling of Gaussian random fields, such as the spectral method where samples are generated in frequency space. A review of such methods can be found in \cite{liu2019advances}. This can allow for Gaussian random fields to be sampled more efficiently than the directly sampling from a large covariance matrix. Second, an analytic approximation of excursion probability, from \cite{Adler:2007-06-12}, can be used to fit a small set of null significance map samples. Finally, given a stationary Gaussian field and a Euclidean parameter space, we have demonstrated that it is possible to directly compute the excursion probability based on the covariance function or template matching template~\cite{Adler:2007-06-12, Worsley:doi:10.1038/jcbfm.1992.127}. These methods can be combined to further reduce the computational cost of estimating the look-elsewhere effect correction at low $p$-values.

We then demonstrate these techniques on 2D and 1D toy problems. The 2D toy problem represents, for example, searches for dark matter using pixel detectors \cite{DAMIC:2013bio, DAMIC:2021crr} and searches for astronomical transients \cite{Andreoni:2019jlb, LSST:2022kad}. 
The 1D toy problem represents searches in a 1D parameter space, such as searches for dark matter using accelerometers \cite{Windchime:2022whs}, sonar \cite{Abraham2019}, and fast radio burst detection \cite{Petroff:2021wug}. Using \(10^5\) samples generated with each method, we show that the look-elsewhere effect corrections derived using toy MC significance maps agree with those sampled from Gaussian random fields, both when the Gaussian random field covariance functions are directly sampled and when the Gaussian random fields are sampled using the spectral method. Finally, a much smaller sample of \(10^3\) null significance maps is used to fit the excursion probability. This analytic fit also agrees with the other approaches, allowing for a greater reduction in computational cost. We also demonstrate that the analytic fit matches the Euler characteristic computed directly using \autoref{eq:euler_characteristic}.

Finally, we have applied these techniques to a \(4^3\) accelerometer array based on the Windchime concept. 
We find that when we require a global significance of \(3\sigma\) the estimated trial factor for such an accelerometer array is \(10^{14}\) for a \(1\unit{s}\) search, and \(10^{22}\) for a \(1\unit{yr}\) search. Taken together, the methods we introduce can help speed up the computation of trial factors in high-dimensional statistical problems, such as that encountered in track finding for Windchime.

\begin{acknowledgments}
We thank Uzu Lim (Oxford University) for extremely helpful discussions about random field theory, especially with regard to the computation of the spectral moments of a Gaussian random field. This work was supported by the U.S. DOE Office of Science, Office of High Energy Physics, QuantISED program (under FWP ERKAP63), and by the U.S. DOE Office of Science, National Quantum Information Science Research Centers, Quantum Science Center.

\end{acknowledgments}


\bibliography{main}

\end{document}